# Un contributo al dibattito sui rapporti tra i cataloghi stellari di Tolomeo e Ipparco

Davide Neri

Liceo Sabin, Bologna (retired)

**Abstract**

The uncertainty about the relationships between the star catalogue contained in the Almagest (books VII-VIII) and the catalogue of Hipparchus dates back to the origins of modern astronomy. The fact that Hipparchus' catalogue was lost and that his observations of a limited number of stars remained has favored the formulation of various hypotheses on the subject. The recent discovery of (at least) a part of Hipparchus' catalogue seems to indicate that there are various concordances between the two catalogues, but also differences sufficient to suggest that Ptolemy's catalogue also derives from sources independent of Hipparchus' catalogue.
This article presents an additional element for the comparison between the two catalogues, based on the diversity of the observation locations of the two astronomers, which provides an argument against the originality of the Ptolemaic catalogue.

**Sommario**

L'incertezza sui rapporti tra il catalogo stellare contenuto nell'*Almagesto* (libri VII-VIII) e il catalogo di Ipparco risale alle origini dell'astronomia moderna. Il fatto che il catalogo di Ipparco fosse perduto e di lui rimanessero dati relativi alle osservazioni di un numero limitato di stelle ha favorito la formulazione di varie ipotesi sull'argomento, che vanno dalla semplice trascrizione (con correzione delle longitudini richiesta dalla precessione degli equinozi) fino alla completa indipendenza. La recente scoperta di (almeno) una parte del catalogo di Ipparco sembra indicare che tra i due cataloghi vi sono varie concordanze, ma anche differenze tali da suggerire che il catalogo di Tolomeo deriva anche da fonti indipendenti dal catalogo di Ipparco.
In questo articolo si presenta un elemento aggiuntivo per il confronto tra i due cataloghi, basato sulla diversità dei luoghi di osservazione dei due astronomi, che fornisce un argomento a sfavore della originalità del catalogo tolemaico.

## La relazione tra i cataloghi di Ipparco e Tolomeo

Tycho Brahe fu il primo a sospettare che il catalogo stellare dell'*Almagesto* di Tolomeo [1] non fosse altro che una trascrizione di quello di Ipparco con l'aggiunta di una correzione (quantitativamente errata) delle longitudini stellari effettuata da Tolomeo per tenere conto della precessione degli equinozi.

Solo nel secolo scorso si è fatta strada l'idea che i due cataloghi siano in realtà indipendenti, sulla base del fatto che il confronto tra i dati superstiti di Ipparco, reperibili nei *Commentari ad Arato* [2], e quelli di Tolomeo non mostrava una differenza meramente riducibile alla correzione sulla longitudine [3,4].

Altri punti di vista si sono collocati tra i due estremi, facendo leva su vari fattori:

– è quasi certo che, mentre nell'*Almagesto* sono contenute le posizioni di circa 1020 stelle, il catalogo di Ipparco non doveva contenerne più di 850 [5,6]: se ciò è vero la corrispondenza può essere solamente parziale, ma resta comunque sconosciuta la provenienza dei dati delle 170 stelle aggiunte da Tolomeo;

– l'analisi statistica degli errori presenti nei due insiemi di dati, pur dimostrando che la relazione tra le coordinate non è perfettamente lineare, come invece dovrebbe essere nel caso della ipotesi di Tycho, rivela la presenza di correlazioni significative, che difficilmente possono essere attribuite al caso [vedi 7, cap. 5];

– inoltre, nei due insiemi di dati vi sono errori di posizione di notevole ampiezza relativi alle stesse stelle e, nel caso di una stella (θ Eridani), si presenta anche un errore grossolano sulla magnitudine: entrambi la considerano di prima grandezza, mentre si tratta di una stella doppia di magnitudine complessiva +3,20 [vedi 7, cap. 3];

– è comunque opinione prevalente tra gli storici che Tolomeo non si è limitato a correggere (in modo errato) i dati di Ipparco, ma ha utilizzato varie fonti nella preparazione del suo catalogo, aggiungendovi un contributo personale.

Pochi anni fa una parte del catalogo di Ipparco è stata rintracciata, per mezzo di analisi multispettrali, in un testo cancellato e sovrascritto nel *Codex Climaci Rescriptus* (un codice del IX secolo d.C. originariamente conservato nel Monastero di Santa Caterina, nel Sinai). I dati recuperati sono compatibili con quelli già disponibili nei *Commentari ad Arato* dello stesso Ipparco. Essi sembrano confermare che l'influenza di Ipparco su Tolomeo può esistere, ma non è completa come speravano i sostenitori della "copiatura". [8]

Secondo quanto affermato dagli studiosi che hanno analizzato il *Codex Climaci Rescriptus* [8], alcune parti del testo sottostante sono irrimediabilmente perdute, e in ogni caso il catalogo "nascosto" sembra incompleto. Si può sperare che altre sue parti si trovino sotto il manoscritto di altri codici conservati nello stesso monastero, ma se non si avrà a disposizione un catalogo stellare completo rimarrà spazio per la formulazione di varie ipotesi alternative sulla in/dipendenza da quest'ultimo del catalogo di Tolomeo.

## I diversi luoghi di osservazione e le loro conseguenze

In questo articolo si vuole presentare un piccolo contributo al problema del rapporto tra i due cataloghi che si basa sui diversi punti di osservazione dei due astronomi: Rodi per Ipparco e Alessandria d'Egitto per Tolomeo.

Rodi si trova a 36°N di latitudine; Alessandria è invece a 31°N. Da ciò consegue che, a causa della sfericità della Terra, i due astronomi erano impossibilitati a osservare una calotta del cielo australe. Tuttavia, essendo Alessandria più a Sud di Rodi, Tolomeo poteva vedere, nell'arco dell'anno, più stelle di quelle che poteva vedere Ipparco. (vedi figura 1)

Un semplice esercizio, che si può

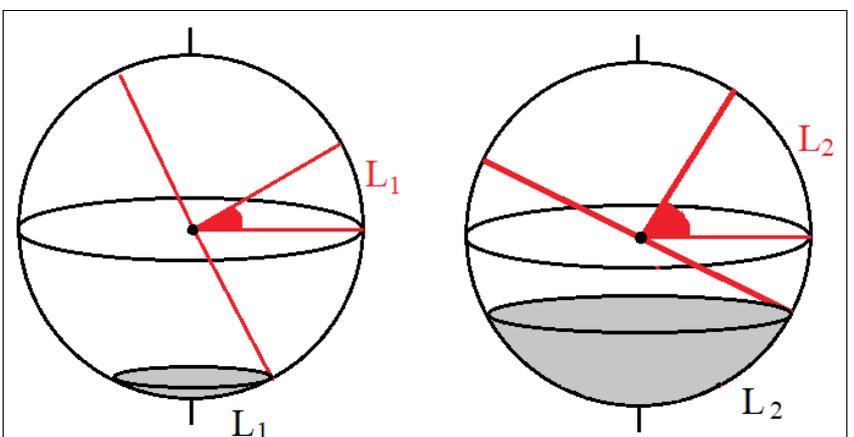

*Figura 1. La regione del cielo australe invisibile per due osservatori che si trovano nell'emisfero Nord si estende dal Polo Sud fino alla codeclinazione corrispondente alla latitudine dell'osservatore.*

effettuare su carta oppure al computer con un elementare software per la gestione di immagini come Paint, permette di stabilire un fatto piuttosto curioso.

Utilizzando una mappa del cielo australe, si può:
- evidenziare quali sono le stelle contenute nel catalogo di Tolomeo;
- individuare dove era posizionato il Polo Sud all'epoca di Ipparco (circa 128 a.C.) e all'epoca di Tolomeo (circa 138 d.C.);
- tracciare il confine dell'area preclusa alle osservazioni di Ipparco, centrata nel Polo Sud calcolato per l'epoca delle sue osservazioni;
- tracciare il confine dell'analoga area per Tolomeo, riferita al Polo Sud del suo tempo.

Se si usa come base una mappa con proiezione azimutale equidistante, le aree di invisibilità sono rappresentate con buona approssimazione da due cerchi. La deformazione che risulta dal tipo di proiezione considerato non è molto importante nella zona interessata, perché in prossimità del polo le ellissi indicatrici di Tissot mostrano una piccola deformazione. A sostegno di questa approssimazione vi è anche una considerazione astronomica che presenterò in seguito.

L'insieme di queste informazioni è riassunto nella seguente figura 2:

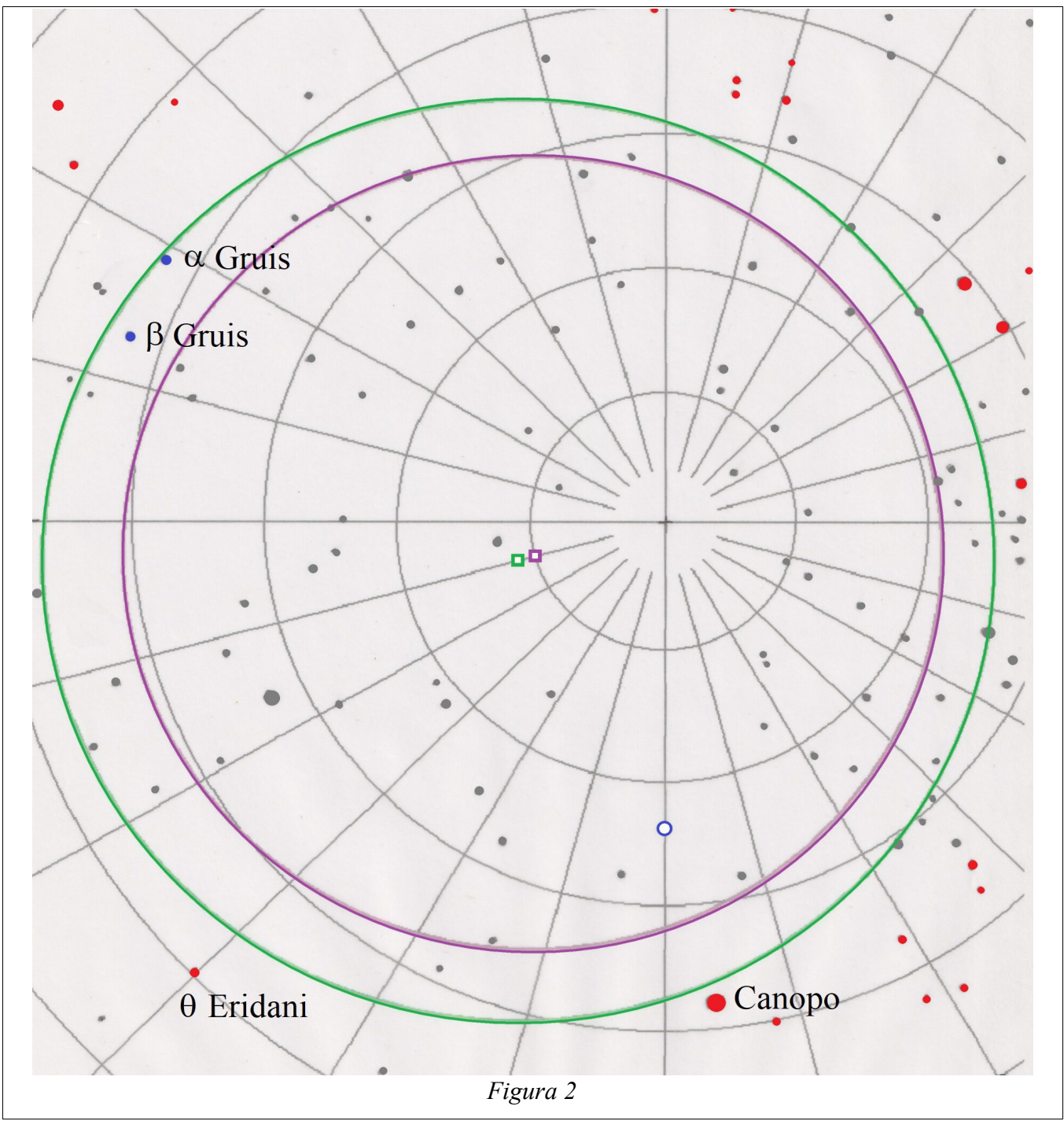

*Figura 2*

La mappa è riferita all'anno 2000.
Il polo dell'Eclittica è rappresentato da un cerchio bianco con bordo blu.
Le stelle che fanno parte del catalogo dell'*Almagesto* sono colorate in rosso.
Il quadrato bianco con bordo verde indica la posizione del Polo all'epoca di Ipparco e la circonferenza verde, centrata in questo Polo, costituisce il limite di visibilità di Ipparco; il raggio della circonferenza (misurato in gradi di declinazione) è 36°.
Analogamente, il quadrato bianco con bordo viola rappresenta il Polo di Tolomeo e la circonferenza viola il suo limite di visibilità. In questo caso il raggio è di 31°, misurati nell'unità della declinazione.
Sulla base della figura 2 si possono fare alcune osservazioni:

1] tutte le stelle del catalogo di Tolomeo sono all'esterno della circonferenza verde. Pertanto nell'*Almagesto* sono catalogate solo stelle che potevano essere viste a Rodi. Non c'è quindi la prova che possa indicare una chiara origine alessandrina;

2] la stella Canopo (α Carinae) è una stella di magnitudine –0,74 che Ipparco cita nei suoi *Commentari*, nonostante essa si levasse a Rodi di un solo grado sopra l'orizzonte (!) e nella mappa risulta molto vicina alla circonferenza verde. Se la deformazione di tale circonferenza fosse molto accentuata (dovendosi trasformare in una ellisse con un semiasse che si allarga nel senso della ascensione retta e non della declinazione) tale stella finirebbe fuori dall'area di visibilità di Ipparco. Ciò significa che la figura rappresenta la situazione reale con una buona approssimazione;

3] nella figura 2 si nota che, nell'area compresa tra la circonferenza verde e quella viola, cioè nell'area che solo Tolomeo poteva osservare, vi sono due stelle rappresentate con dischi blu (α e β Gruis) di magnitudine apparente +1,74 e +2,13, che tuttavia non figurano nel catalogo dell'*Almagesto*. Si tratta di stelle che sono, rispettivamente, più luminose di α e β Ursae Majoris. Ipparco aveva osservato una stella molto meno luminosa, la già citata θ Eridani, con un'elevazione sull'orizzonte molto simile a quelle che avevano α e β Gruis per Tolomeo (vedi ancora la figura 2).

## Le posizioni delle stelle mancanti

Utilizzando il programma *Almagest Stars* [9] è possibile visualizzare il cielo e le coordinate valide all'epoca di Tolomeo anche per le stelle che non sono state inserite nell'*Almagesto*, ricavate dallo *Yale Bright Star Catalogue* e ricalcolate per la data del 20 luglio 137.
La tabella presenta le coordinate eclittiche di α e β Gruis (e quelle equatoriali corrispondenti, ricalcolate):

|  | longitudine λ | latitudine β | asc. retta α | declinazione δ |
| --- | --- | --- | --- | --- |
| α Gruis | 289°51' | –32°39' | 299°11' | –54°06' |
| β Gruis | 296°12' | –35°15' | 309°10' | –55°11' |

Le declinazioni così ottenute concordano molto bene con quanto si poteva dedurre qualitativamente dalla figura 2. I risultati confermano che Tolomeo poteva vedere le due stelle, distanti dal Polo rispettivamente 36° e 35°, perché si elevavano sull'orizzonte di Alessandria rispettivamente di 5° e 4°.

## Conclusione

La presenza di α e β Gruis nel catalogo dell'*Almagesto* avrebbe rafforzato la tesi della indipendenza di Tolomeo da Ipparco. La loro assenza sembra indicare che Tolomeo non fece una ricognizione sistematica del cielo a lui visibile e amplifica l'impressione che il suo sia stato più che altro un lavoro di assemblaggio di fonti precedenti. Pur non potendo modificare molto i termini della discussione, questa lacuna appare, col senno di poi, un'occasione mancata per avere chiarezza sul rapporto tra i due cataloghi.

## Bibliografia

[1] Ptolemy, *Almagest* (traduzione e commento di G.J. Toomer), Princeton University Press, 1998.

[2] K. Manitius, *Hipparchi in Arati et Eudoxi Phaenomena Commentariorum libri tres*, Teubner, 1894.

[3] F. Boll, *Die Sternenkataloge des Hipparch und des Ptolemaio*, Bibliotheca Mathematica, 3. Folge, Bd. 2, pp. 185-195.

[4] H. Vogt, *Versuch einer Wiederherstellung von Hipparchus Fixsternverzeichnis*, Astr. Nachr., 224

(1925) no. 5354-5355, cols. 2-48.

[5] O. Neugebauer, A *History of Ancient Mathematical Astronomy*, Springer, 1975, vol. 1, pp. 274-298.

[6] G.J. Toomer, *Hipparchus*, in: C.C. Gillispie (a cura di), *Dictionary of Scientific Biography*, Supplement I, C. Scribner's sons, 1978, pp. 207-224 (in particolare vedi pp. 216-218).

[7] G. Grasshoff, *The history of Ptolemy's star catalogue*, Springer, 1990.

[8] V. Gysembergh, P.J. Williams e E. Zingg, *New evidence for Hipparchus' Star Catalogue revealed by multispectral imaging*, Journal for the History of Astronomy, 53 (2022) 383-393.

[9] Il programma *Almagest Stars* è reperibile all'indirizzo http://www.etwright.org/astro/almagest.html.